# Excitonic Josephson effect induced by interlayer tunneling currents: Robust evidence for exciton condensation

Ya-Fen Hsu[1,2,*] and Jung-Jung Su[2,†]

[1]*Physics Division, National Center for Theoretical Science, Hsinchu, 30013, Taiwan*
[2]*Department of Electrophysics, National Chiao Tung University, Hsinchu 300, Taiwan*
(Dated: January 24, 2022)

The Josephson effect can be regarded as a striking manifestation of exciton condensation. It has been suggested to tune the condensate phase of bilayer excitons by applying interlayer tunneling current. A poorly-understood phenomenon observed by Huang *et al* [Phys. Rev. Lett. **109**, 156802 (2012)] demonstrates that the critical values of the interlayer tunneling current at either edge can be controlled by passing a second interlayer tunneling current at the other edge. We successfully attribute this novel coupling to excitonic Josephson effect induced by tunneling-current generated relative phases and indicate that Huang's experiment is very robust evidence for exciton condensation. We furthermore make a new proposal: there exists a critical Josephson current beyond which Josephson coupling collapses and external currents prefer to convert into edge-state currents to meet Huang's observation at the second tunneling current above$_\pm$ 16nA — a sudden decoupling of two edges accompanied by a large interedge voltage. Also, we make a feasible suggestion to detect Josephson current by measuring induced magnetic field of a ring-shaped excitonic Josephson junction. This work not only solves the long-standing uncertainty regarding the identification of exciton condensation but also opens a new direction to explore exciton condensation.

*Introduction.*— The Bardeen-Cooper-Schrieffer like physics of exciton condensates (electron-hole pairs) has stimulated considerable theoretical and experimental activity in quantum Hall bilayers[1, 2]: a few fascinating transport anomalies, e.g., Josephson-like tunneling, quantized Hall drag, and vanishing counterflow resistance were reported. In the standard Hall bar geometry, however, the edge-state currents lead to the ambiguity for identifying exciton condensation[3]. Later, the Corbino geometry[4–6] was experimentally employed to rule out the contribution of edge-state currents but rigorous theoretical works really in the context of the Corbino geometry are still lacking so that the existing experimental works stay in naive elucidation concerning observed "phenomena". This leaves the demand for the exact identification of exciton condensation. More recently, Huang *et al*[7, 8] conducted a Corbino experiment demonstrating that the critical values of external interlayer tunneling current measured at either edge are correlated with the second tunneling current exerted on the other edge, where the critical values occur as the interlayer voltage suddenly emerges. Amazingly, in this letter, we show this novel coupling of two edges is originated from excitonic Josephson effect[9–11] induced by interlayer tunneling currents and Huang's experiment can be regarded as robust evidence for exciton condensation since the Josephson effect is well-known as a striking manifestation of coherent condensation[12].

In Huang's experiment[7], the function of external currents $[J_{L(R),in}, J_{L(R),out}]$ is equivalent to applying tunneling currents $J_{tL(R)}$ (coherent part associated with Andreev reflection) as well as to injecting edge-state quasiparticle currents $J_{eL(R)}$ (incoherent part) [see Fig. 1 and Ref. [13]]. The Andreev reflection[5, 14] propagates excitons at the edge of a quantum Hall bilayer by pairing injected electrons with reflected ones in different layers, making a equal number of electrons flow into one layer and out of the other layer synchronously like external tunneling current. The remnant unpaired electrons would turn into incoherent edge-state currents. Deduced from Park and Das Sarma's ingenious idea[15]: the condensate phase of bilayer excitons can be tuned by externally applying tunneling current, $J_{tL}$ and $J_{tR}$ twist the condensate phase of two edges to embed the relative phases— the integral ingredient for Josephson effect. Huang's device can be regarded as a excitonic Josephson junction with a superfliud barrier[12]. The resultant Josephson current $J_s$ serves as the link of two edge tunneling currents and therefore leads to the observed novel coupling[7].

Along this line, our calculation yields the critical values of $J_{tR}$ versus $J_{tL}$ curves that behave in full agreement with the experimental results[7], confirming our theoretical explanation as well as exciton condensation in Huang's device. To suggest the detection of Josephson current, we also calculate the induced magnetic field in a ring-shaped excitonic Josephson junction and derive a measurable magnitude ∼10pT for a practicable device. Recently, double-graphene-layer structures — one of the most promising candidates for exciton condensation — attract enormous attention[19, 20]. The present work opens up a new opportunity to exactly identify exciton condensation in these novel systems.

*Theoretical method.*— The quantum Hall bilayer can be mapped onto a pseudospin-1/2 ferromagnet of the normalized pseudospin orientation $m(x) = (m_\perp \cos\varphi, m_\perp \sin\varphi, m_z)$ in which pseudospin up (down) corresponds to the top- (bottom-) layer quantum degree of freedom while $\varphi$ ($m_z$) represents the condensate phase (layer charge imbalance)[16, 17]. The external tunneling current $J_t$ has two actions: (i) to make electrons flow out of the top layer and (ii) to make electrons flow into the bottom layer. They respectively decrease and increase the $z$ component of total pseudospin ($nW\delta L dm_z/dt$) by ($J_t W\delta L/e$)$\times$(1) and by ($J_t W\delta L/e$)$\times$(−1), where $W\delta L$ is the cross area of external tunneling current (see Fig. 1). Under the effect of external tunneling currents at two edges, we are supposed to modify the $z$ component of the Landau-Lifshitz-Gilbert (LLG) equation that describes the pseudospin dynamics of quantum Hall bilayers[11, 18] as follows:

$$\frac{dm_z}{dt} = -\frac{2\rho_s}{n\hbar}m_\perp^2 \nabla^2 \varphi + \frac{\Delta_t}{\hbar}m_\perp \sin\varphi - \frac{2J_t}{ne} + am_\perp^2\frac{d\varphi}{dt},$$
$$J_t = J_{tL}\Theta(x+L/2)\Theta(L/2-\delta L - x)$$
$$+ J_{tR}\Theta(L/2-x)\Theta(x-L/2+\delta L). \quad (1)$$

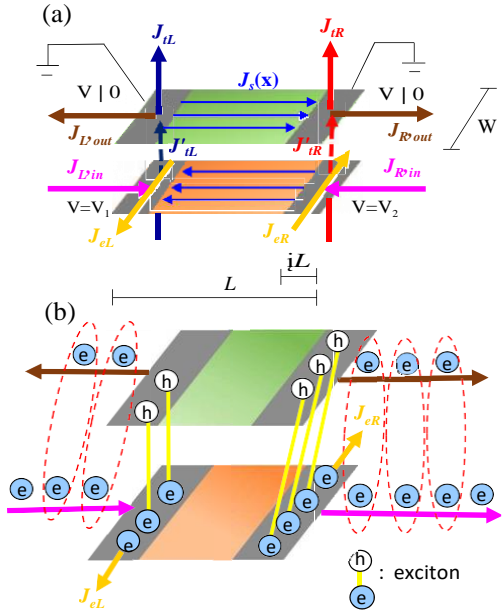

FIG. 1. (a) Schematic layout of Huang's device — an excitonic Josephson junction: two edge condensates of phases controlled by external tunneling currents are sandwiched by a superfluid barrier, being a type of Josephson junction as a superconducting analogy demonstrating in Ref. [12]. (b) Decomposition of external currents: Andreev-reflection coherent part (from paired electrons) and remnant incoherent part (from unpaired electrons).

Here the parameters $\rho_s$, $n$, $\Delta_t$, $\alpha$ and $L$ denote the superfluid density, pseudospin density, single-particle tunneling energy, Gilbert damping coefficient and junction length, respectively.

In Huang's experiment, some part of $J_{L(R),in}$ is coherent with $J_{L(R),out}$ via Andreev reflection, forming a current pair that acts like external tunneling current. We therefore can explore Huang's experiment by the above tunneling-current model including $L = 0.6\lambda$ and $\delta L \sim 200$nm, where $\lambda = (2\rho_s/n\Delta_t)^{1/2}$[13, 21, 22]. We numerically solve the modified LLG equation[11, 18] and determine the critical values of $J_{tR}$ by finding the upper and lower boundaries at which the static solution of $m_z$ starts to become nonzero[23]. The remains of $J_{L(R),in}$ would be transformed into incoherent edge-state currents staying in the bottom layer and leads to a voltage difference ($V_2 - V_1$) across two edges so that the effective external tunneling current should be measured in the top layer (see Fig. 1 and Ref. [13]).

*Coupling of two edge tunneling currents.—* It is widely observed that the interlayer voltage suddenly occurs when external tunneling current attains certain critical values, namely, Josephson-like tunneling[2, 6, 9]. Huang's experiment[7] reports the critical values: the upper and lower $I_c$ are correlated with the second tunneling current exerted on the other edge [see Fig. 2(a)]. Our calculation within $|J_{tL}| \leq 10.551 J_{t0}$ (10.551$J_{t0}$ is the intrinsic value for critical tunneling currents and $J_{t0} = en\Delta_t/2\hbar$ is the unit for tunneling currents) produces the same characteristics presented in the experimental curves of the critical currents versus the second tunneling current [see Figs. 2(b)-(c)][24]: (i) the magnitudes of critical currents decreases (increases) with increasing the second tunneling current of parallel (antiparallel) polarity and (ii) the slopes for the calculated and measured curves show similar trend. Characteristic (i) can be comprehended through Figs. 2(d)-(f) (which are deduced from our calculation[25]):

In the static limit and the coherent tunneling regime ($m_z = 0$), Eq. (1) is reduced to

$$J_t = -\nabla \cdot J_s + J_t',$$
$$J_s = e\rho_s \nabla\varphi/\hbar,$$
$$J_t' = ne\Delta_t \sin\varphi/2\hbar, \quad (2)$$

where $J_s$ and $J_t'$ denote Josephson current and internal tunneling current. The underlying physics behind this equation is the Kirchhoff's law: the total current inflow and outflow are equal at any node to maintain zero net charge everywhere. Therefore for parallel polarity [see Fig. 2(d)], Josephson current going along opposite directions are collected together in the bulk and then transformed into large internal tunneling current. In comparison to the case without the second tunneling current applied [see Fig. 2(e)], for antiparallel polarity [see Fig. 2(f)], more electron/hole charges can be dispersed into the left half of the junction by large Josephson current so as to lower internal tunneling current since the remnant electrons and holes must annihilate with each other by internal tunneling current. On the other hand, the critical points of $J_{tR}$ occur as $\varphi \approx \pm \pi/2$ ($J_t' \approx \pm J_{t0}$), where $J_t'$ reaches its maximum[26]. It is therefore easier/harder (needs to input smaller/larger $J_{tR}$) to reach critical points for parallel/antiparallel polarity.

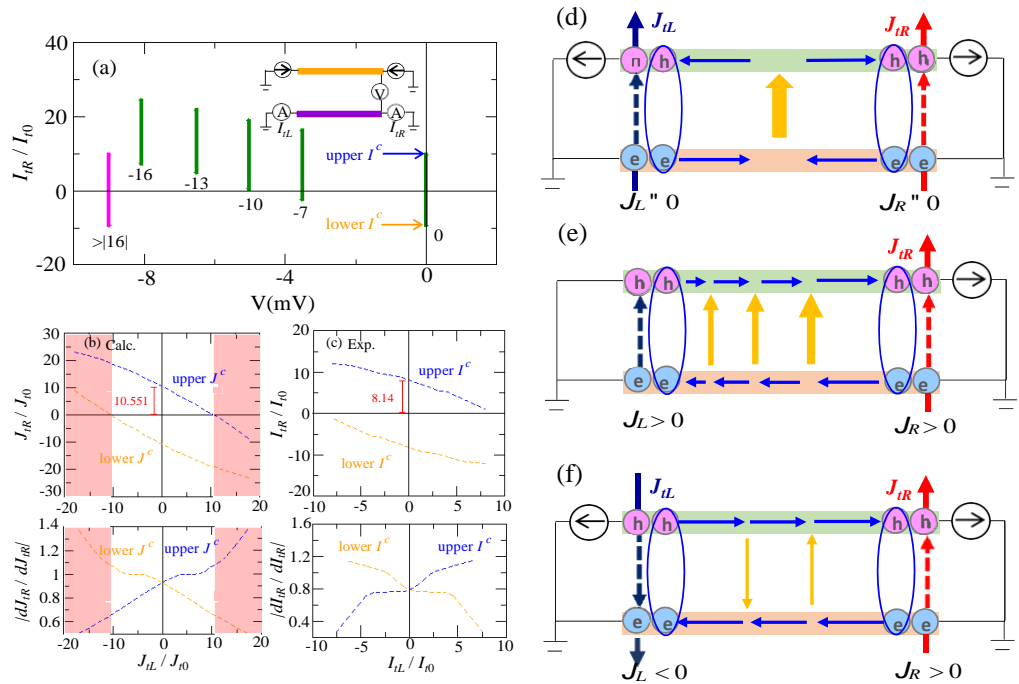

FIG. 2. (a) Measured Josephson-like $I$-$V$ characteristic for different values of a second tunneling current passed at the other edges, reproduced from Fig. 3(a) of Ref. [7]: the numbers below the traces labels the value of the second tunneling current $I_{tL}/I_{t0}$ and the $I$-$V$ curves are offset by $(I_{tL}/2I_{t0})$mV, where $I_{t0}$ = 1nA[24]. The inset shows the measurement configuration. (b) Calculated upper and lower critical currents (top) and corresponding slopes (bottom) as a function of the second tunneling current: the unit for tunneling currents reads $J_{t0} = en\Delta_t/2\hbar$. (c) Experimental results corresponding to (b), reproduced from Fig. 2(b) of Ref. [7]. In (b) and (c), the red numbers denote the intrinsic values for critical tunneling currents and the shaded region of (b) is with $J_{tL}$ exceeding the intrinsic critical current. (d), (e), and (f) depict current flows for parallel polarity, without the second tunneling current applied at the other edge and for antiparallel polarity, in which the yellow vertical (blue horizontal) arrows denote internal tunneling current (Josephson current) and their width (length) indicate the magnitudes of internal tunneling current (Josephson current).

*Exotic decoupling beyond the intrinsic critical tunneling current.*— Another surprise found by Huang's experiment[7] is when increasing the second tunneling current beyond 16nA, the coupling of two edges would suddenly disappear [see Fig. 2(a)]: this occurs at the region of $I_{tL} >$ the system's intrinsic critical tunneling current (8.14nA) and requires a positive (negative) $I_{tR}$ prior to the application of the negative (positive) $I_{tL}$. In our opinion, possibly, like conventional Josephson effect[12], there exists a critical value for Josephson current beyond which Josephson effect would collapse. As is well known, the conventional junction becomes normal conducting once exceeding the critical Josephson current. Analogously but subtle differently, our junction bulk is supposed to go insulating since it is constructed from a quantum Hall bilayer. Then external currents would prefer to convert into edge-state currents so as to engender a voltage across two edges, meeting Huang's observation[7]: the sudden disappearance of coupling of two edges is accompanied by the emergence of a large interedge voltage.

We identify the state as being static and of zero $m_z$ by the average time-varying rate $\overline{dm/dt} \to 0$ as $t \to \infty$ and the average in-plane pseudospin component $m_\perp = 1$, respectively. With a moderate tuning ($\Delta J_{tR} = 0.001 J_{t0}$), we find there exists no static $m_z = 0$ solution when $|J_{tL}| \gtrsim 25 J_{t0}$ [see Figs. (a)-(c)]: the solution is dynamic as $J_{tR} < 14.343 J_{t0}$ & $J_{tR} > 35.653 J_{t0}$ and static but $m_z \neq 0$ elsewhere for $J_{tL} = -25 J_{t0}$. We therefore regard $J_{tL} = \pm 25 J_{t0}$ as the point where the critical Josephson current occurs and this value is a bit larger than twice the intrinsic critical tunneling current close to the experimental result: $16/8.14 \sim 2$. Furthermore, we analyze the critical value for Josephson current. Fig. 2(d) shows the spatial extrema of Josephson current $J_s^{\text{extre}}$ is larger at the upper and lower critical points and Fig. 2(e) furthermore shows $J_s^{\text{extre}}$ at the two critical points depends only on the difference of external tunneling currents $\Delta J_t$ [27]. We find $J_s^{\text{extre}}$ always increases with increasing $\Delta J_t$ and beyond the intrinsic critical tunneling currents, $\Delta J_t$ is raised by $J_{tL}$. Increasing $J_{tL}$ raises $J_s^{\text{extre}}$ towards its critical value so as to cause exotic decoupling and our calculation shows this critical value is $1.4 J_{s0}$ ($J_{s0} = e\rho_s/\hbar\lambda$ is the unit for Josephson current and this critical value occurs at $\Delta J_t \sim \pm 50 J_{t0}$).

*Experimental suggestion for detecting Josephson current.*— To suggest the detection of Josephson cur-rent, we consider a Josephson junction bent into a ring shape, as shown in Fig. 4(a). Such geometry will generate circular Josephson current so as to induce magnetic field and can be simply fabricated by cutting a Corbino disk with a small slit. We are interested in two candidates: Huang's Corbino ($\lambda \quad 2\pi R_{\min} < 2\pi R_{\max}$[7]) and graphene Corbino ($2\pi R_{\min} < 2\pi R_{\max} < \lambda$[28]). We calculate the induced magnetic field $B_J$ by the Biot-Savart Law[29] and find the two short junctions[22] display (nearly) linear curves of $B_J$ versus $\Delta J_t$ [see Figs. 4(b) and 4(c)[30]]. Specifically, for Huang's Corbino, $B_J$ is on the order of $10^{-11}/\lambda$ tesla ($\lambda$ in unit of meter) and the order reads $\sim 10$pT that is measurable by using scanning superconducting quantum interference device (SQUID)[31]. However, the contribution of edge-state currents is unavoidable when using side electrodes. We have suggested some method to solve this problem in Sec. SM.V.C of supplemental material.

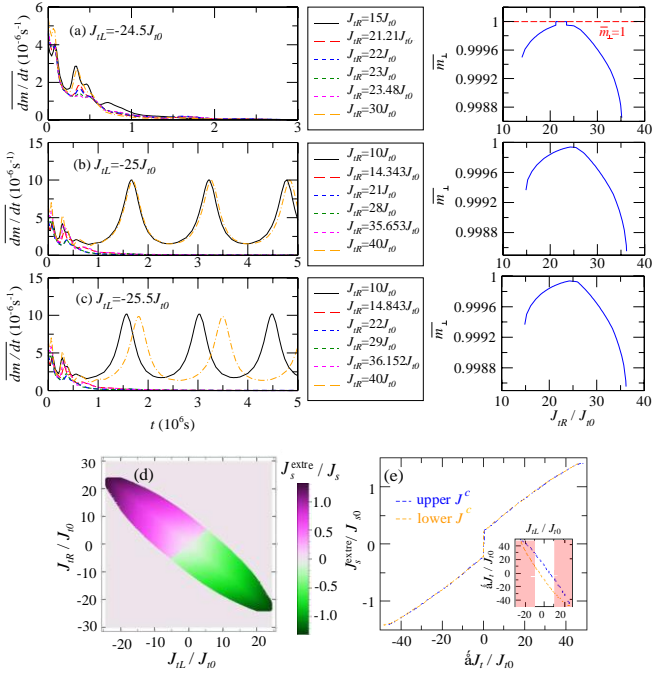

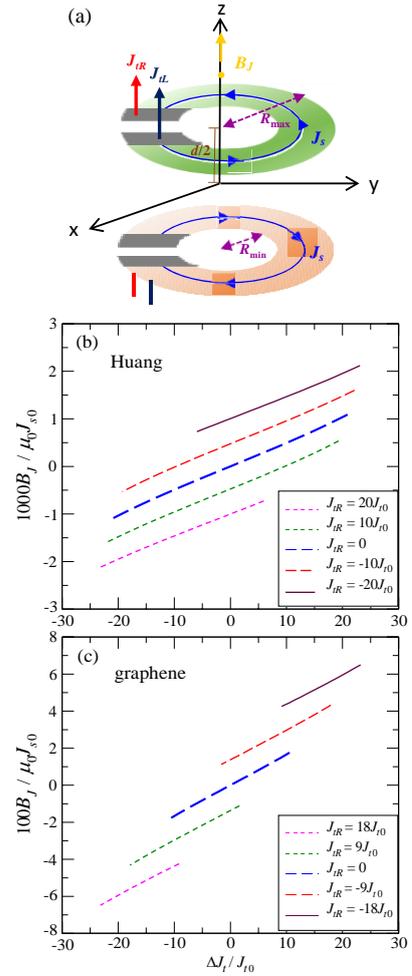

FIG. 3. (a), (b) and (c) are the time-varying rate ($\overline{dm/dt}$) and the in-plane component ($\overline{m}_\perp$) of the normalized pseudospin on average per discretization pixel for the second tunneling current $J_{tL} = -24.5 J_{t0}$, $J_{tL} = -25 J_{t0}$ and $J_{tL} = -25.5 J_{t0}$, respectively. (d) and (e) are spatial extrema of Josephson current in the plane of tunneling currents ($J_{tL}, J_{tR}$) and as a function of $\Delta J_t = J_{tR} - J_{tL}$, respectively. The insert shows how $\Delta J_t$ corresponds to $J_{tL}$ and the shaded region is with $J_{tL}$ exceeding the intrinsic critical current. The current units for tunneling currents and Josephson current read $J_{t0} = en\Delta_t/2\hbar$ and $J_{s0} = e\rho_s/\hbar\lambda$.

*Conclusion.*— In conclusion, we have built up the physical mechanism for Huang's experiment[7]— the excitonic Josephson effect induced by tunneling currents, which may be by far the most robust evidence for exciton condensation. The present model even may be extended to explain another fascinating experiment[32] that still lacks theoretical understanding. The only constraints are Landau quantization (broadly speaking, topological quantization[33]) and bilayer structure and therefore it is applicable to a wide range of systems[34–39]. Notably, two additional ingredients — interlayer tunneling and quantum Hall nature — create richer physics relative to conventional Josephson effect[12], e.g., a unique dependence on the polarity of tunneling currents and a sudden decoupling of two edge condensates, and the high controllability by tunneling currents can bring new energy to technological progress. The tunneling-current induced excitonic Josephson effect indeed offers a new route to explore exciton condensation and we believe it is hopeful to stimulate enormous research activity in the future.

FIG. 4. (a) Schematic layout of a ring-shaped excitonic Josephson junction: the induced magnetic field $B_J$ is observed at the $z$-axis. (b) and (c) are the induced magnetic field due to circular Josephson current at $z = 0.11\lambda$ as a function of the difference of two external tunneling currents $\Delta J_t = J_{tR} - J_{tL}$ for Huang's Corbino[7]: $R_{\min} = 0.36\lambda$ and $R_{\max} = 0.95\lambda$ and for graphene Corbino[28]: $R_{\min} = 0.095\lambda$ and $R_{\max} = 0.15\lambda$, respectively. The curves are offset by their respective $J_{tL}$[30] and $J_{t0}$ ($J_{s0}$) is defined as in Fig. 3. The interlayer separation reads $d = 1.6\ell_B$, where $\ell_B$ is the magnetic length of $\sim 20$nm.

*Acknowledgements.*—We are grateful to W. Dietsche, A. H. MacDonald, B. Rosenstein, Jheng-Cyuan Lin, Sing-Lin Wu, Chien-Ming Tu and Ming-Chien Hsu for valuable discussion. This work were financially supported by Ministry of Science and Technology and National Center for Theoretical Sciences of Taiwan.

---

# Supplemental Material for "Excitonic Josephson effect induced by tunneling current: Robust evidence for exciton condensation"

Ya-Fen Hsu

*Physics Division,*
*National Center for Theoretical Science, Hsinchu, 30013,*
*Taiwan*
*Department of Electrophysics,*
*National Chiao Tung University, Hsinchu 300,*
*Taiwan*

Jung-Jung Su

*Department of Electrophysics,*
*National Chiao Tung University, Hsinchu 300,*
*Taiwan*

In this supplemental material, we show how Huang's experimental device corresponds to an excitonic Josephson junction with relative phases induced by tunneling currents, the analytical calculation of induced magnetic field in ring-shaped excitonic Josephson junctions and how to rule out the contribution of edge-state current in the magnetic-field measurement as well as any other complementary information.

**CONTENTS**

## SM.I. CORRESPONDENCE OF HUANG'S EXPERIMENTAL DEVICE TO AN EXCITONIC JOSEPHSON JUNCTION

Here we illustrate in detail why Huang's device(1) can be viewed as an excitonic Josephson junction through Fig. SM. 1. In Fig. SM. 1(a), the specific contacts numbers 1-5 and 1*-5* denote the position of contacts at the top and bottom layer, respectively. To correspond with Huang's experiment, we suppose the input current are sent from contact 2* to 2 and from 4* to 4 and both two ends of the top layer (contacts 2* and 4*) are grounded as depicted in Fig. SM. 1(b). This supposition just exchanges the roles of the top and bottom layers in Huang's experiment and does not distort the truth of its physics. We are only interested in the transport behavior within the blue box. As have shown in the main body of the manuscript, the function of external currents can be decomposed into two contributions: (i) interlayer-tunneling-current-like coherent part and (ii) edge-state-current incoherent part. The contributions of Part (ii) can be formulated as pseudospin-transfer torque(2) which serves pseudospin conservation as quasiparticles go through noncollinear magnetic domains. In quantum Hall effect, edge-state current must be driven by in-plane electric field(3). Since in absence of in-plane electric field in the top layer (this is because two ends are grounded there), the edge-state current flows only in the bottom layer (with collinear magnetization $m_z = -1$) and contributes nothing to pseudospin-transfer torque. Under the consideration of pseudospin conservation, we just need to take Part (i) into account.

As have shown in the main body of the manuscript, the effect of external interlayer tunneling current $J_t$ can be formulated as

$$\frac{dm_z}{dt} = -\frac{2\rho_s}{n\hbar}m_\perp^2 \nabla^2\varphi + \frac{\Delta_t}{\hbar}m_\perp \sin\varphi - \frac{2J_t}{ne} + \alpha m_\perp^2 \frac{d\varphi}{dt}. \quad \text{(SM1)}$$

Similarly, by considering pseudospin conservation, the effect of coherent part of external side currents $J_{out}$ decreases the $z$-component of total pseudospin $(nW\delta L dm_z/dt)$ by $(2J_{out}W/e)$:

$$\frac{dm_z}{dt} = -\frac{2\rho_s}{n\hbar}m_\perp^2 \nabla^2\varphi + \frac{\Delta_t}{\hbar}m_\perp \sin\varphi - \frac{2J_{out}}{ne\delta L} + \alpha m_\perp^2 \frac{d\varphi}{dt}, \quad \text{(SM2)}$$

where $J_{out} = J_{L,out}\Theta(x + L/2)\Theta(L/2 - \delta L - x) + J_{R,out}\Theta(L/2 - x)\Theta(x - L/2 + \delta L)$ (for the definitions of $J_{L(R),out}$, see Fig. 1 in the main body of the manuscript). It is apparent by comparing Eqs. (SM1) and (SM2) that the external currents $J_{L(R),out}$ creates effective tunneling currents $J_{L(R),out}/\delta L$. The exciton coherence is maintained over one correlation length $\xi$ so a effective $\delta L$ of $\sim \xi$ can model Huang's experiment. The typical value of $\xi$ reads $\sim 200$nm(4).

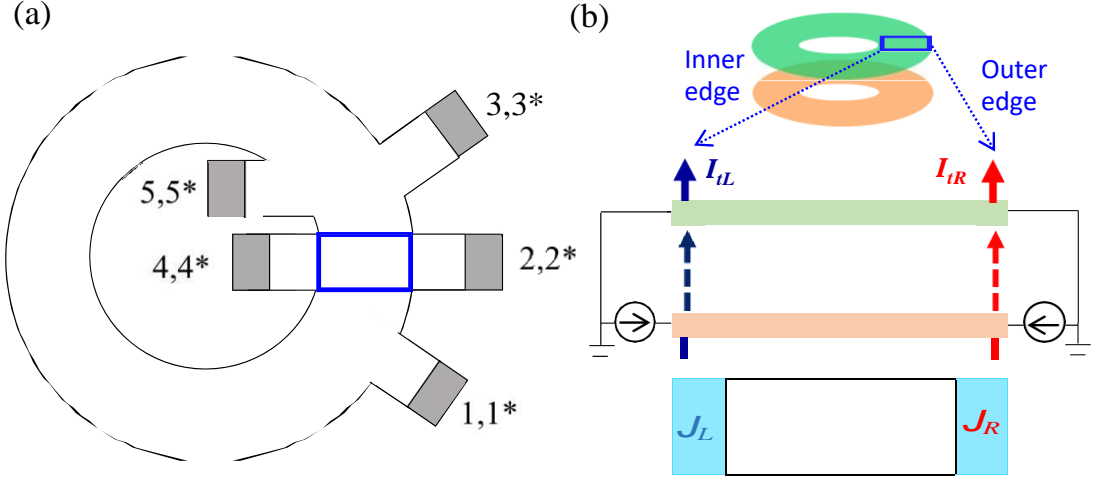

Fig. SM. 1: (a) Schematic layout of the Corbino geometry for each layer. (b) The excitonic Josephson junction cut from the Corbino bilayer.

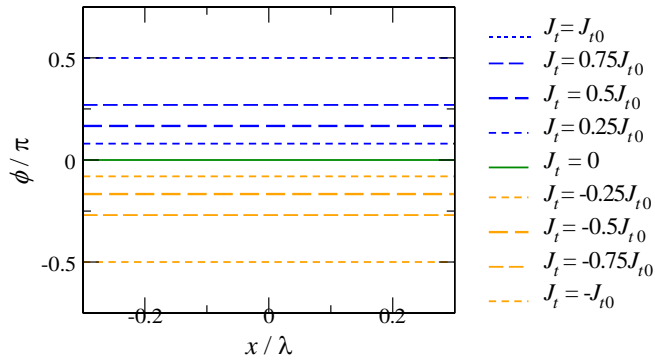

Fig. SM. 2: The phase distribution for applying constant tunneling current $J_t$ over the whole quantum Hall bilayer. Here $J_{t0} = en\Delta_t/2\mathsf{n}$.

The LLG equation would relax the pseudospin into a static configuration. In Fig. SM. 2, we show the finally obtained phase distribution for the quantum Hall bilayer with uniform tunneling current applied to confirm the suggestion of Ref. (5). We find the phase can vary from $-\pi/2 \sim \pi/2$ with applying tunneling current. That is to say, we can apply different tunneling currents to two edges so as to create the relative phase between them. Two edge condensates are sandwiched by a superfluid barrier, being a type of Josephson junctions(6). We thus deduce that the Huang's experiment (1) can be analyzed by studying an excitonic Josephson junction with junction length$\sim (R_{out} - R_{in})$, where $R_{out}$ and $R_{in}$ are outer and inner radius [see Fig. SM. 1(b)].

## SM.II. THE ORIGINAL CALCULATED DATA GRAPHS FOR SCHEMATIC DIAGRAMS OF CURRENT FLOWS

The Josephson current $J_s$ and internal tunneling current $J_t^{'}$ can be derived from the static solution of the phase $\varphi$ that is obtained by solving the LLG equation: $J_s = e\rho_s \nabla\varphi/\mathsf{n}$ and $J_t^{'} = ne\Delta_t \sin\varphi/2\mathsf{n}$. Fig. SM. 3 shows our calculation results for the distribution of the phase and its spin-offs — the distributions of Josephson current and internal tunneling current for different polarities. The phase profiles for parallel polarity [see Fig. SM. 3(a)], without the second tunneling current applied [see Fig. SM. 3(b)], and for antiparallel polarity [see Fig. SM. 3(c)] are parabolic, half-parabolic, and nearly linear, respectively. This indicates the first one has Josephson current following in opposite directions while the later two similarly have Josephson current following along the same direction. But, it is different between the later two: without the second tunneling current applied, Josephson current will go to zero near the left edge while for antiparallel polarity, Josephson current will remain finite and almost constant. These inferences can also be clearly seen in Fig. SM. 3(d). From the magnitude of $\varphi$, we also can infer that applying the second tunneling current with parallel (antiparallel) polarity will raise (lower) $J_t^{'}$ as shown in Fig. SM. 3(e). Therefore, these calculation results can be schematically depicted as Figs. 2(d)-(f) in the main body of the manuscript.

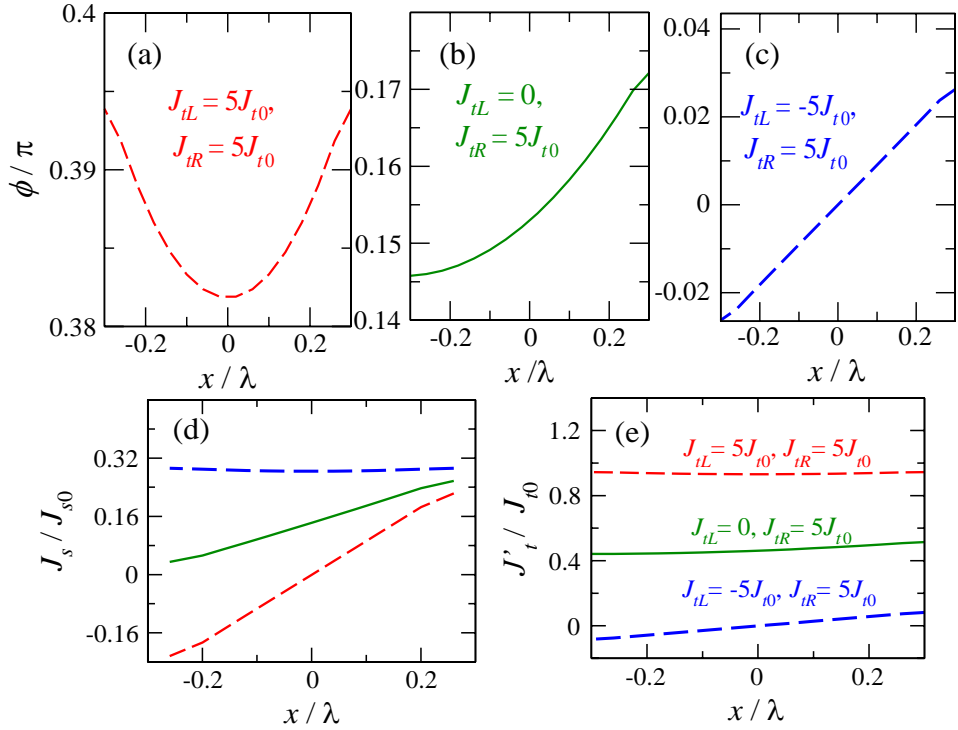

Fig. SM. 3: The phase distributions (a) for parallel polarity: $J_{tL} = 5J_{t0}$, $J_{tR} = 5J_{t0}$, (b) without the second tunneling current applied at the other edge: $J_{tL} = 5J_{t0}$, $J_{tR} = 0$, and (c) for antiparallel polarity: $J_{tL} = -5J_{t0}$, $J_{tR} = 5J_{t0}$ and their corresponding (d) Josephson current and (e) internal tunneling current. Here $J_{s0} = e\rho_s/\hbar\lambda$ and $J_{t0} = en\Delta_t/2\hbar$.

## SM.III. CONDITIONS FOR CRITICAL POINTS OF $J_{tR}$

In this section, we identify where the critical points of $J_{tR}$ occur in Fig. SM. 4. We find the critical points always occur at $\varphi \approx \pm\pi/2$ ($J_t^{'} \approx \pm J_{t0}$), where $J_t^{'}$ reaches its maximun since $J_t^{'} = J_{t0}\sin\varphi$ but are not specific to $J_s$.

## SM.V. CALCULATION OF INDUCED MAGNETIC FIELD IN RING-SHAPED EXCITONIC JOSEPHSON JUNCTIONS

### A. The induced magnetic field due to circular Josephson current

The Corbino can be divided into a set of rings with radius ranging from $R_{\min}$ to $R_{\max}$. A single ring of the specific radius $r$ can be viewed as a bent Josephson junction with the junction length $L = 2\pi r$. We calculate the phase distribution of the minimum ring with $L = 2\pi R_{\min}$ by the theoretical method illustrated in the main body of the manuscript and then take azimuthal symmetry into account to obtain the phase distribution of the whole Corbino. Similarly, the critical values of external tunneling current are determined by $L = 2\pi R_{\min}$ and the Josephson current is calculated by $J_s = e\rho_s \nabla \varphi/\hbar$. Finally, the induced magnetic field $B_J$ is calculated by the Biot-Savart Law as follows:

$$B_J(z) = \frac{\mu_0 \langle J_s(R_{\min}, \theta)\rangle_\theta \, zdR_{\min}}{2} \left[ \frac{1}{(R_{\min}^2 + z^2)^{3/2}} - \frac{1}{(R_{\max}^2 + z^2)^{3/2}} \right] \quad (SM3)$$

where $d$ is the interlayer separation, $z$ is the distance above the center of the bilayer, and $\langle \cdot \rangle_\theta$ is the average over the angular axis of polar coordinate.

### B. The without-offset version of subfigures in Fig. 4

It can be seen from Fig. SM. 5, the curves for different $J_{tR}$ are almost overlapped. By comparing Figs. SM. 5(a) and 5(b), we can find enlarging the Corbino size would give rise to a slight deviation. The induced magnetic field is almost a single value function of $\Delta J_t$ and unrelated to the individual values of external tunneling currents for short junctions which are actually in strong Josephson coupling. That is to say, the Josephson effect drives the system into being bulk-dominant.

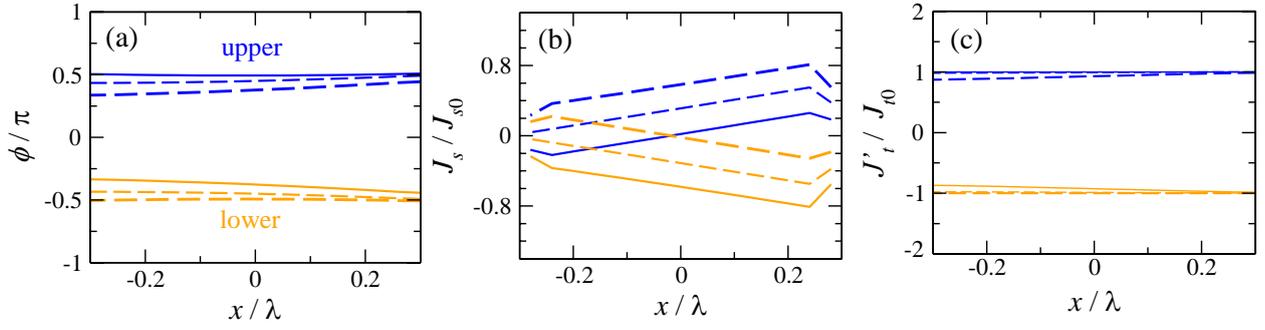

Fig. SM. 4: The distributions of (a) the phase, (b) Josephson current, and (c) internal tunneling current for upper and lower critical points of $J_{tR}$ occurring at $J_{tL} = 5J_{t0}$ (solid line), $J_{tL} = 0$ (dashed line), and $J_{tL} = -5J_{t0}$ (dotted line). The upper [lower] critical points $(J_{tR}, J_{tL})$ read $(5.669J_{t0}, 5J_{t0})$ [$(-14.904J_{t0}, 5J_{t0})$], $(10.551J_{t0}, 0)$ [$(-10.551J_{t0}, 0)$], and $(14.904J_{t0}, -5J_{t0})$ [$(-5.669J_{t0}, -5J_{t0})$]. Here $J_{s0} = e\rho_s/\text{n}\lambda$ and $J_{t0} = en\Delta_t/2\text{n}$.

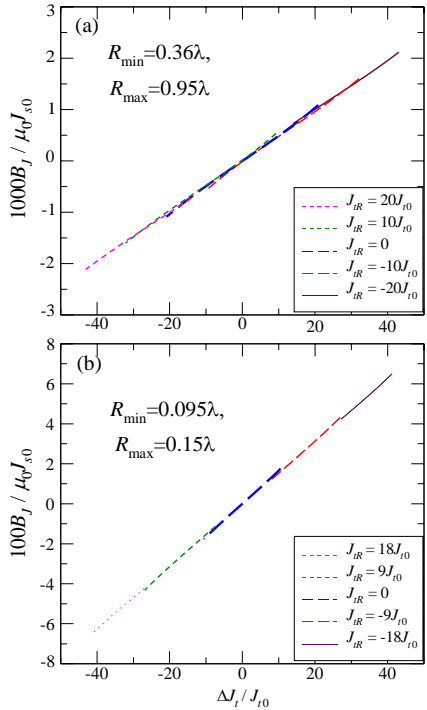

Fig. SM. 5: (a) The without-offset version for Fig. 4(b). (b) The without-offset version for Fig. 4(c). Here $B_J$ and $\Delta J_t$ denote the induced magnetic field and the difference of two external tunneling currents, respectively. Here $J_{s0} = e\rho_s/\text{n}\lambda$ and $J_{t0} = en\Delta_t/2\text{n}$.

### C. The methods to rule out the contribution of edge-state current in the magnetic-field measurement

As stated in the main body of the manuscript, it is difficult to avoid edge-state current if we use the side electrodes. The usage of top- and back- electrode pairs can solve this problem but the fabrication of back electrode is challenging due to the existence of substrate. Another method is to take the contribution of edge-state current out in measurement. By using the Biot-Savart Law, the induced magnetic field due to edge-state currents $J_{eL(R)}$ can be calculated as

$$B_e(z) = \frac{\mu_0}{2}\left[\frac{J_{eL}R_{min}^2}{(R_{min}^2 + z^2)^{3/2}} - \frac{J_{eR}R_{max}^2}{(R_{max}^2 + z^2)^{3/2}}\right]. \quad \text{(SM4)}$$

$B_e(z)$ and $B_J(z)$ [see Eq. (SM3)] are symmetrical and antisymmetrical about the $xy$ plane, respectively. Therefore, if the experimenters measure the total magnetic field $B = B_J + B_e$ at two mirror points and then sum the two measured values, the contribution of edge-state current can be separated out, i.e., $[B(z) + B(-z)]/2 = B_e(z)$. The pure $B_J(z)$ can furthermore be obtained by subtracting the measured $B(z)$ by $B_e(z)$.